\documentclass[fleqn,usenatbib]{mnras}

\usepackage{newtxtext,newtxmath}
\usepackage{multirow}
\usepackage{appendix}
\usepackage[T1]{fontenc}

\DeclareRobustCommand{\VAN}[3]{#2}
\let\VANthebibliography\thebibliography
\def\thebibliography{\DeclareRobustCommand{\VAN}[3]{##3}\VANthebibliography}

\usepackage{graphicx}	% Including figure files
\usepackage{amsmath}	% Advanced maths commands

\graphicspath{ {./Thesis/Thesis_Figures/} {./MG_MSW/} {./SG_MSW/} }
\title{On Atomic Line Opacities for Modeling Astrophysical Radiative Transfer}

\author[Morag]{
Jonathan Morag\thanks{E-mail: jmorag88@gmail.com}
\\
$^{1}$Weizmann Institute of Science, Rehovot, Israel\\
}

\pubyear{2024}

\begin{document}
\label{firstpage}
\pagerange{\pageref{firstpage}--\pageref{lastpage}}
\maketitle

\begin{abstract}
In astrophysics, atomic transition line opacity is a primary source of uncertainty, leading to orders of magnitude discrepancy in theoretical calculations of radiative transfer in the literature. Much of this uncertainty is dominated by the inability to resolve the lines in frequency, leading to the use of approximate frequency-averaged treatments, often employing the `line-expansion formalism'. In this short paper we assess the usage of this formalism, specifically the prominent Eastman \& Pinto 1993 formula (hereafter EP93).
As a case study, we reproduce EP93 opacities from the commonly-used STELLA simulations in order to highlight the orders of magnitude effect due to the choice of line treatment.
We show that the widely used EP93 expansion opacity substantially underestimates photon emissivity and reprocessing rates, even when it correctly captures photon mean-free-paths.
We also highlight the importance of introducing micro-plasma electron excitation level cutoffs in the equation of state (EOS) for calculating opacity.

An alternative method for calculating emissivity is based on a simple frequency-binned average of the lines. We introduce a physically-motivated modification to this method that leads to a minor reduction in the calculated opacity. A fully-consistent coarse-frequency solution does not currently exist for line modeling.

Finally, we describe new features in our updated publicly available high-resolution frequency-dependent opacity table.

\end{abstract}

\begin{keywords}
radiation: dynamics – supernovae: general
\end{keywords}

%%%%%%%%%%%%%%%%%%%%%%%%%%%%%%%%%%%%%%%%%%%%%%%%%%

%%%%%%%%%%%%%%%%% BODY OF PAPER %%%%%%%%%%%%%%%%%%
\defcitealias{morag_shock_2024}{M24}
\defcitealias{morag_frequency_2023}{M23}
\defcitealias{blinnikov_comparative_1998}{B98}
\defcitealias{eastman_spectrum_1993}{EP93}
\section{Introduction}

Calculating radiative transfer in high-energy astrophysical contexts often requires input of the photon to plasma interaction cross-section ("opacity"), which introduces considerable theoretical uncertainty.
A primary challenge involves the implementation of sharp atomic transition lines, whose widths and separations in wavelength can be many orders of magnitude smaller than the resolution currently available in simulation, and whose strengths can be orders of magnitude above the scattering opacity. Due to these and additional challenges (e.g. incomplete corresponding laboratory measurements, common uncontrolled assumptions such as plasma local thermal equilibrium, and complex plasma microsphysics), a fully self-consistent calculation of the "bound-bound" line interaction is not currently available.

In expanding supersonic flows, the presence of a thick forest of lines can significantly reduce the propagating photon's mean-free-path as it Doppler shifts in frequency. This effect is often included in line treatments that derive a coarse frequency-averaged approximation of the opacity for use in radiative transfer \citep{karp_opacity_1977,friend_stellar_1983,eastman_spectrum_1993,blinnikov_correct_1997}. These "expansion opacity" formalisms are all based on similar assumptions (see \S~\ref{sec: bound-bound opacity}) and have been shown to be in general agreement with each other \citep{castor_radiation_2007,potashov_opacity_2021}, especially in the limits of all-weak or all-strong lines\footnote{\citet{blinnikov_correct_1997} actually provides a mono-chromatic description for the bound-bound opacity, and has been shown, after coarse frequency averaging, to be in reasonable agreement with \citet{eastman_spectrum_1993}, at least relative to the orders of magnitude disagreements we report in this letter.}.
Of these methods, the formalism of \citet[][hereafter \citetalias{eastman_spectrum_1993}]{eastman_spectrum_1993} has been commonly used in  the community \citep[e.g.][]{blinnikov_comparative_1998,tominaga_shock_2011,forster_delay_2018,ben_nasr_atomic_2023,gallego_statistical_2024},
including in Monte-Carlo simulations \citep{kasen_time-dependent_2006,kawaguchi_diversity_2020,barnes_kilonovae_2021,domoto_lanthanide_2022,bulla_critical_2023}\footnote{Monte-Carlo methods have an intuitive advantage for line forests but face a similar challenge in resolution for describing the emissivity. Other approaches to the problem include using steady-state methods, low optical depths and/or fully incorporating at least a subset of the involved atomic lines without use of the expansion formalism \citep{lucy_monte_2002,lucy_monte_2003,kromer_time-dependent_2009}.}.

The STELLA code \citep{blinnikov_comparative_1998,tominaga_shock_2011,kozyreva_shock_2020} has been used extensively in the literature to calculate the spectral energy density (SED) from transients in high-energy astrophysics. It employs a 1-dimensional radiative transfer `multi-group' treatment, where the photons are binned into frequency groups and radiative transfer for each bin is solved separately. Bound-bound opacity is calculated for each average bin using the \citetalias{eastman_spectrum_1993} treatment based on experimentally verified atomic line lists by \citet{kurucz_atomic_1995}. 
The STELLA SED results for core-collapse supernovae shock cooling emission were compared in the literature to a similar multigroup code employing Kurucz lines \citep[][hereafter \citetalias{morag_shock_2024}]{morag_shock_2024}. The comparison yielded an orders of magnitude disagreement in the SED in certain wavelengths, and this was shown to be due to different respective choices of line treatment.

In this short letter we show that the choice of line treatment is of primary importance in calculating the line opacity. We do this by reproducing the STELLA opacities given in \citet[][hereafter \citetalias{blinnikov_comparative_1998}]{blinnikov_comparative_1998}, and then showing that a change to only the line treatment can yield orders of magnitude difference. We use this comparison to assess the validity of the commonly employed \citet[][hereafter \citetalias{eastman_spectrum_1993}]{eastman_spectrum_1993} prescription. We then describe a minor physics-based modification to the frequency-averaged emission calculation in a way that accounts for line-expansion effects. This paper is written as follows. We reproduce the \citetalias{blinnikov_comparative_1998} photo-ionization "bound-free" opacity in \S~\ref{sec: bound-free EOS}, and bound-bound opacity in \S~\ref{sec: bound-bound opacity}. In \S~\ref{sec: alternative approach} we outline the physics-based modification.
In \S~\ref{sec: Summary Discussion} we discuss and summarize, and also announce useful updates to our publicly available-frequency dependent opacity table, given in \citetalias{morag_frequency_2023}.

\section{Bound-Free Opacity and Equation of State (EOS)}
\label{sec: bound-free EOS}

We first attempt to reproduce the STELLA bound-free opacity. In wavelengths where bound-bound lines are present, the bound-free opacity should have only secondary effect. However, since it is simpler to calculate, such a comparison can be useful for isolating other aspects of the calculation including the underlying equation of state (EOS) of the species in the plasma.
Computing the EOS requires the use of a limiting physical cutoff for the allowed excited bound electron states due to micro-interactions between nearby species in the plasma. Without such a cutoff, the atomic partition function diverges. In our opacity table \citepalias{morag_frequency_2023} we address this effect in Hydrogen by adopting a prescription from \citet{hummer_equation_1988}, which forbids highly excited states due to the presence of nearby ions in the plasma. We use these states to solve the Saha equation self-consistently assuming LTE.

\begin{figure}
    \centering
    \includegraphics[width=\columnwidth]{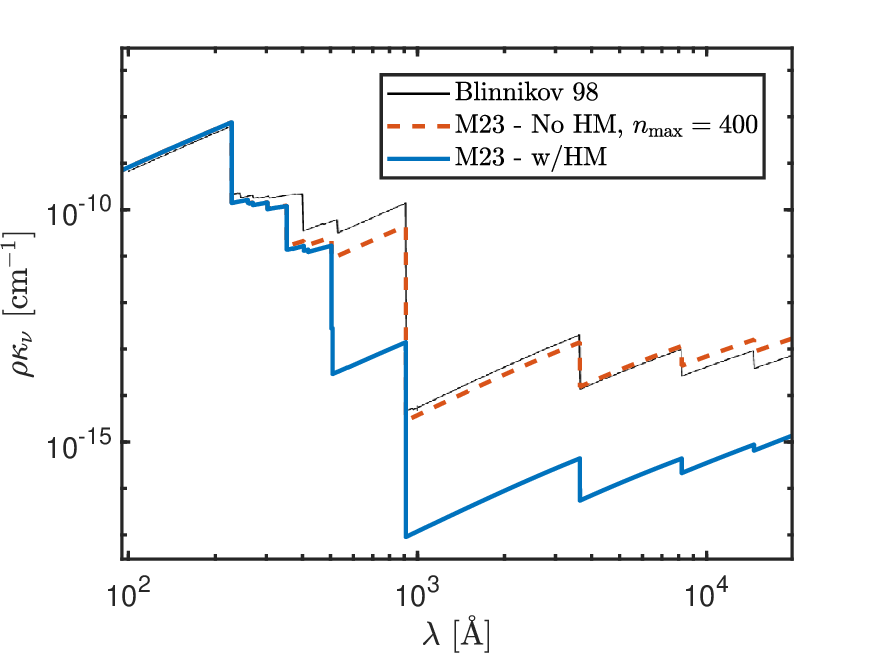}
    \caption{
    	Our imitation of the bound-free opacity in \citetalias{blinnikov_comparative_1998} (black line) for the example case of $\rho=10^{-13} ~ \rm ~cm^{-3}$, $T=15,000$ K for a solar mixture. In red dashed-line, we show the result when the \citet{hummer_equation_1988} factor is not included, and we limit the Hydrogen partition function to $n_{\rm max}=400$, finding reasonable agreement with \citetalias{blinnikov_comparative_1998}. In blue solid lines we show the result of a converged H partition function, representative of what we insert into the simulations in \citetalias{morag_shock_2024}. The difference in the opacities can be orders of magnitude in the H photoinization opacity (He photoionization is less affected).
    }
    \label{fig: bf opacity}
\end{figure}

In fig. \ref{fig: bf opacity} we compare only bound-free opacity $\rho\kappa_\nu$ from \citetalias{blinnikov_comparative_1998} fig. 1 at top (here in black line) to that produced by our table (blue line). There is an orders of magnitude difference between the two tables for Hydrogen photoionization peaks ($\lambda>500$ \AA).  Since the photo-ionization for Hydrogen is given by simple analytic relations, and since the He cross-sections ($\lambda\le500$ \AA) are in reasonable agreement for the two formulations, the difference between the opacities is very likely a result of different implementations of the H equation of state.  Specifically, there is a likely difference in the ionization level and the population of electrons at each excited atomic level.

We are able to reproduce the \citetalias{blinnikov_comparative_1998} opacity to a factor of a few when we don't implement the Hummer Mihalas factor in Hydrogen, and when we arbitrarily cut off the number of Hydrogen levels to $n_{\rm max}=400$ (in this case, Hummer Mihalas should greatly restrict the presence of electrons above $n\sim 10's - 100$).
As EOS calculations tend to be complex, the exact reason for the discrepancy is unknown to us, and we do not know which implementation is more correct. We note however, that that the \citetalias{morag_frequency_2023} table was shown to agree with the publicly available TOPS table \citep{colgan_new_2016} with regards to the bound-free opacity to $\sim10\%$ \citepalias{morag_shock_2024}.

The presence of the strong Hydrogen photo-ionization cross-section in \citet{blinnikov_comparative_1998} likely accounts for the sharp photo-ionization cutoff observed in the SED in \citet{tominaga_shock_2011}, which uses STELLA. Due to our much lower bound-free opacity, and different implementation of atomic transition lines (see below) we do not observe such a cutoff when using our \citetalias{morag_frequency_2023} table (see \citetalias{morag_shock_2024}). 

We also note that the introduction of the Hummer Mihalas factor can either increase $\kappa_\nu$ or decrease it, depending on implementation and whether or not it is employed in the Saha equation to decide ionization, or in the bound-free and bound-bound opacities to determine the bound electron excitation populations. In general, some excitation level cutoff and test of convergence should be included in both.

\section{Bound-Bound Opacity}
\label{sec: bound-bound opacity}

\begin{figure}
    \centering
    \includegraphics[width=\columnwidth]{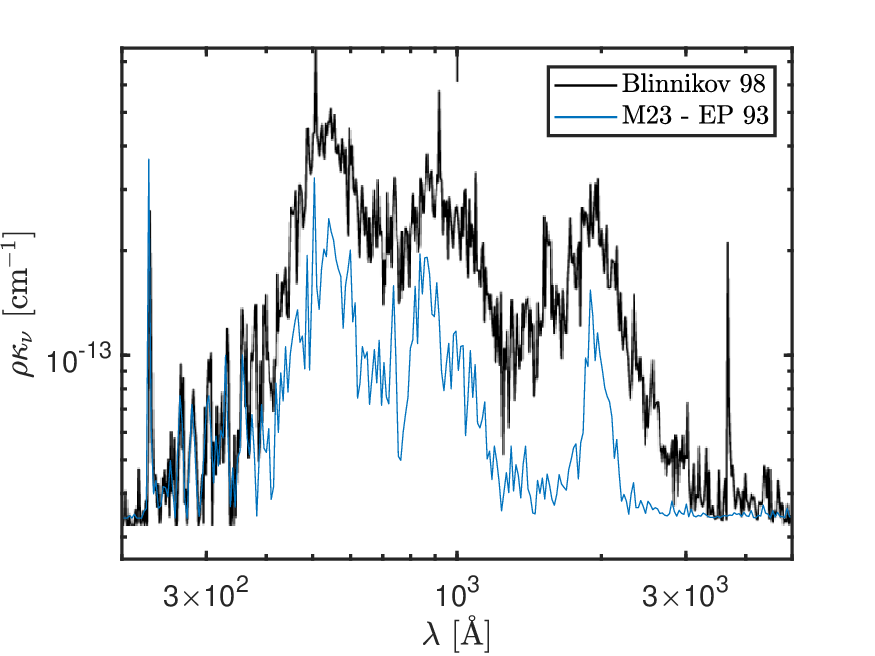}
    \caption{
    Bound-bound opacity example from \citetalias{blinnikov_comparative_1998} fig. 1, compared with our reproduction using \citetalias{morag_frequency_2023} modified to employ the \citetalias{eastman_spectrum_1993} prescription. We  find good agreement to a factor of a few or better. The plasma parameters are $\rho=10^{-13} ~ \rm ~cm^{-3}$, $T=15,000$ K, $t_{\rm exp}=15$ days. Similarly to \citetalias{blinnikov_comparative_1998}, we use a coarse frequency grid with $\Delta\nu/\nu\sim0.01$. The lines that are in excellent agreement in the range 200 \AA ~ $ < \lambda <$ 400 \AA ~ are dominated by a Helium line ($\lambda=227$\AA) and a set of Oxygen lines.
    }
    \label{fig: bb opacity higher-res}
\end{figure}

In fig. \ref{fig: bb opacity higher-res} we attempt to reproduce an example STELLA bound-bound opacity, taken from fig. 1 -center- in \citetalias{blinnikov_comparative_1998}.
Both opacities shown use the frequency-averaged line strengths according to 
\citetalias{eastman_spectrum_1993}, as defined below. The plasma microphysics in our reproduction (electron and ion populations, as well as equation of state) are based on methods in \citetalias{morag_frequency_2023}, sampled in a frequency grid of $(\Delta\nu/\nu)_{\rm i}\sim0.01$, where the i subscript denotes a binned frequency group.
We find agreement with STELLA in the line peaks to tens of percents, while the agreement in the continuum is nominally a factor of a few and up to nearly an order of magnitude. Correspondence is not exact, due to differences in inserted microphysics, but is much closer than the several orders of magnitude discrepancies due to choice of line treatment shown in the subsequent figure.
% Normally, emission and absorption opacities from our high frequency resolution \citetalias{morag_frequency_2023} opacity table are calculated using a frequency average across each bin, $\left<\kappa_\nu\right>_{\rm i}$.

\citetalias{eastman_spectrum_1993} describes the mean free-path of a photon in a line forest as it Doppler shifts in frequency in the expanding flow. It is given by
\begin{equation}
	\chi_{\rm exp,i} = \rho \kappa_{\rm exp,i}= (\nu/\Delta\nu)_{\rm i} (ct_{\rm exp})^{-1} \sum_l [1-\exp(-\tau_l) ],
	\label{eq: Eastman Pinto 93}
\end{equation}
where the sum $l$ is performed over all lines within the frequency bin i $(\nu,
\nu+\Delta\nu$) containing many lines. The line Sobolev optical depth $\tau_l$ is given by
\begin{equation}
   \tau_l=ct_{\rm exp} [1-\exp(-h\nu_l/T) ]  \sigma_l n_{e,jl} / \nu_l, \quad \sigma_l=\pi \left(e^2 /m_{\rm e}c \right) f_l.
   \label{eq: tau sobolev}
\end{equation}
Here $m_{\rm e}$ is the electron mass, $\nu_l$ and $f_l$ are the line frequency and oscillator strength,
$n_{e,l}$ is the bound-electron number density in the excited state corresponding to the lower energy level of the atomic transition line. The expansion time $t_{\rm exp}$ is a placeholder for the local velocity shear $\sim(d v / d r)^{-1}\sim (v/r)^{-1}$, equivalent everywhere to a single value (the dynamical time $t$) in the case of freely coasting spherical ejecta.
 
\begin{figure}
    \centering
    \includegraphics[width=\columnwidth]{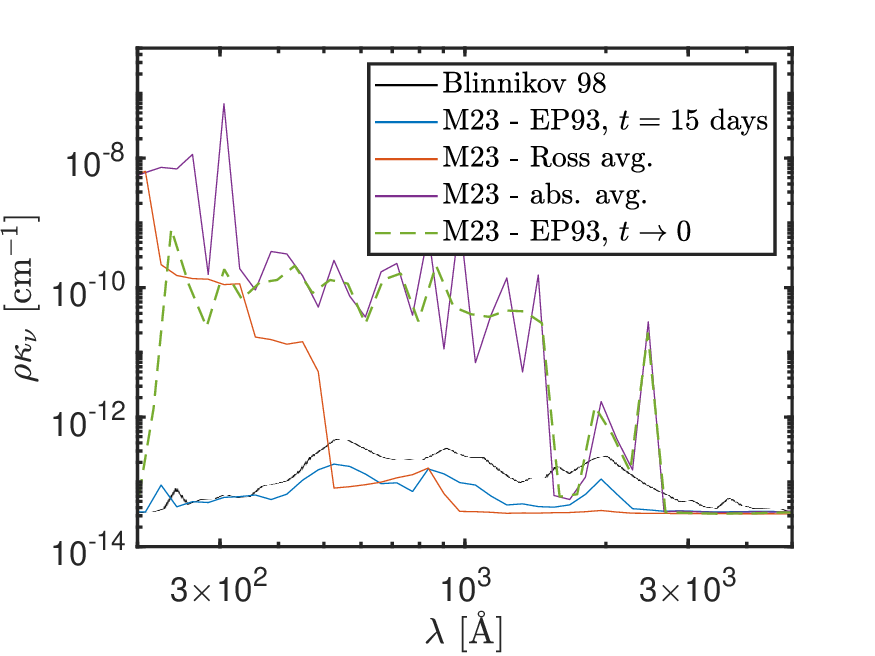}
    \caption{
    Same as fig. \ref{fig: bb opacity Blin vs ours}, at lower frequency resolution ($\Delta\nu/\nu\sim0.1$).
    For ease of comparison, the $\kappa_{\rm es}$ baseline has been added to the average and \citetalias{eastman_spectrum_1993} opacities, despite not normally being added to the emission / absorption terms. The Ross mean opacity also includes bound-free opacities (unlike other opacities in the figure), as these cannot be separated in a harmonic average.
    }
    \label{fig: bb opacity Blin vs ours}
\end{figure}

In fig. \ref{fig: bb opacity Blin vs ours} we show the same comparison but now at a lower frequency resolution ($[\Delta\nu/\nu]_{\rm i}\sim0.1$), similarly to \citetalias{blinnikov_comparative_1998} fig. 1 at bottom. Here we zoom out and add a comparison to two `static' opacities (not incorporating Doppler expansion) at the same density and temperature. Namely, we show both the frequency bin averaged $\left<\kappa_\nu\right>_{\rm i}$ and the bin Rosseland mean $\kappa_{\rm R,i}$, as employed in \citetalias{morag_shock_2024}.
Our static $\kappa_{\rm R}$, which is used to determine radiative transfer (diffusion), is insensitive to the presence of lines. It does not exhibit large deviations from the electron scattering opacity $\kappa_{\rm es}$, and therefore is similar in this case to \citetalias{eastman_spectrum_1993} (not including bound-free opacity  - see fig. \ref{fig: bb opacity Blin vs ours} caption).
On the other hand, $\left<\kappa_\nu\right>_{\rm i}$, which is used in the emission / absorption term, is higher than the \citetalias{eastman_spectrum_1993} result by several orders of magnitude.

We also show in fig. \ref{fig: bb opacity Blin vs ours} that in the early limit, $t_{\rm exp}\to0$, the expansion opacity approaches $\left<\kappa_\nu\right>_{\rm i}$. The behavior in this limit can be deduced analytically from eqs. \ref{eq: Eastman Pinto 93} and \ref{eq: tau sobolev}.
As $t_{\rm exp}$ increases, the rate of photon frequency shift due to velocity shear decreases. Therefore, $\tau_l$ increases linearly with time as photons passing through the line spend longer in resonance. Meanwhile, photons also spend longer traveling in between lines, hence the $(ct_{\rm exp})^{-1}$ factor in eq. \ref{eq: Eastman Pinto 93}. Since the contribution of each strong line is counted as at most $[1-\exp(-\tau_l)]\to1$, the net effect is a reduction of the opacity over time as more lines become stronger and saturate (i.e. $\kappa_{\rm abs}\to0$ as $t_{\rm exp}\to\infty$).
For the ($\rho,T$) parameter choice in this example, the \citetalias{eastman_spectrum_1993} opacity is already lower from the static result by up to an order of magnitude when $t_{\rm exp}=1$ hr. Consequently the effect of lines when using \citetalias{eastman_spectrum_1993} during shock-cooling is very weak.

\section{A physically motivated cutoff to line emissivity}
\label{sec: alternative approach}

As discussed above, \citetalias{eastman_spectrum_1993} should underestimate the emissivity, while the bin-averaged opacity $\langle\kappa_\nu\rangle_i$ may yield an overestimation, especially for strong lines (see discussion in \citetalias{morag_shock_2024}). In this section we partially bridge the gap in emissivity, introducing a correction to the averaging method that includes expansion effects. The modification should be more accurate than the simple averaging method, though its impact on optically thick flows are expected to be moderate.

Consider a line in a static medium. The presence of the line should quickly thermalize the local photon intensity $I_\nu\to B_\nu$\footnote{We assume a thermally emitting plasma with source function $B_\nu(T)$, though these arguments can be generalized to an emitter with a general source function $S_\nu$.} within a time-scale ~$(\rho c\kappa_\nu(\nu_l))^{-1}$, where $\kappa_\nu(\nu_l)$ is the opacity near the peak of the line.  Once the local photon intensity at resonance matches the source function $B_\nu$, the net photon production/absorption from the line will be zero.

In an expanding medium, the photon frequency sweep rate is given by $\partial \nu / \partial t \sim \nu / t_{exp}$.
Assume momentarily that the photons being swept into a strong line are weak in intensity ($I_\nu\ll B_\nu$).
Inside the resonance region, $I_\nu\sim B_\nu$ will be maintained locally, and therefore the maximum rate of \textit{net} photon production -in units of intensity per unit time- will be capped by $B_\nu (\nu / t_{exp})$. i.e., new net photons will be produced at most at the rate that photons will be swept away. Likewise, if the field incoming into the line is strong ($I_\nu\gg B_\nu$), the maximum absorption rate will be given by $I_\nu (\nu / t_{exp})$. i.e. net absorption will only occur at maximum at the rate at which photons are being swept \textit{into} the resonance region.
This reasoning suggests a modification to the line strength, based on the limit beyond which photons will not be produced or absorbed. It is given by
\begin{equation}
\kappa_{l,exp}=\min[\kappa_l, (\rho c t_{exp})^{-1}]
\end{equation}
where $\kappa_l$ is a quantity proportional to line strength that is useful for line expansion. It is defined by $\kappa_\nu\equiv\kappa_l \nu_l \delta(\nu-\nu_l)$.
Line wings should not be modified when using the limit.

One can also assume for simplicity that the peak opacity of an atomic line is approximately $\kappa_\nu(\nu_l) \sim \kappa_l \nu_l/\Delta \nu_l$, where $\Delta \nu_l$ is the line-width (including e.g. thermal broadening, but not including line expansion by Doppler shear). Then one can also derive the above result by equating the thermalization timescale with the timescale for sweeping the photons through the line resonance region, given by ($\Delta \nu_l/\nu_l)\times t_{exp}$.

\begin{figure}
    \centering
    \includegraphics[width=\columnwidth]{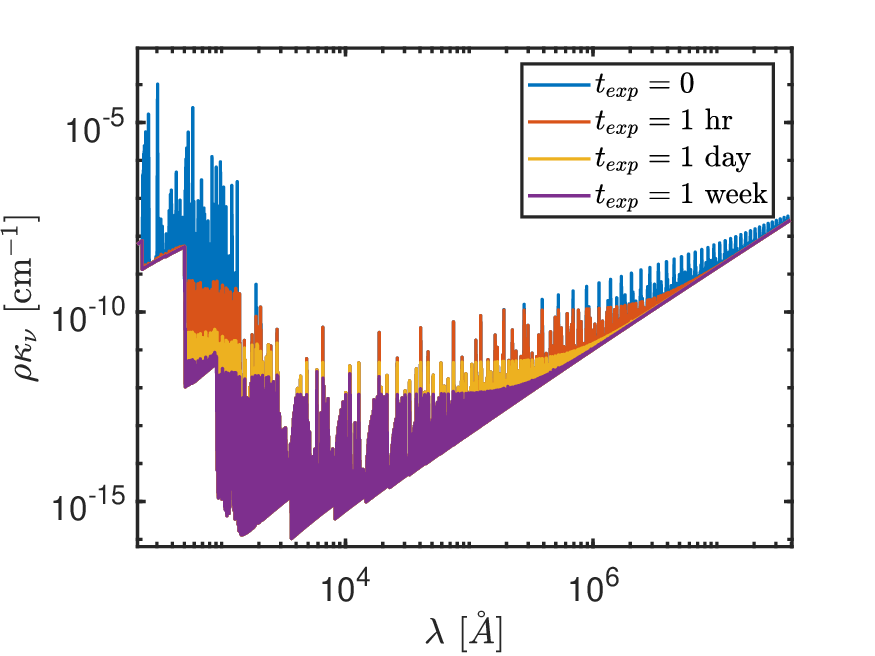}
    \caption{Example high resolution opacity for a solar mixture at density and temperature $\rho=10^{-13} \, \rm g \, cm^{-3}$, $T=1 \, \rm eV$, showing the effect of the proposed expansion limit. As $t_{exp}$ increases,  the maximum line strength decreases, as the \textit{net} photon production rate depends on the rate at which photons are swept in frequency out of the line resonance region. This high-resolution opacity can be averaged for use in coarse multi-group expansion methods. It was created using the publicly available \citet{morag_frequency_2023}, which now includes the expansion limit. }
    \label{fig:example opacity t_exp limit}
\end{figure}

In fig. \ref{fig:example opacity t_exp limit} we show an example of how the proposed limit would affect the \textit{high-resolution} frequency dependent opacity.
In the example shown, with temperatures of 1 eV, the Planck peak (around which photon production tends to be maximal) would still be deep in a line forest even during $t_{exp}=1$ week. A full reassessment of published transient predictions using this prescription would require dedicated radiative-transfer calculations and is left to future work. However, we note that our previous numerical calculations in \citetalias{morag_shock_2024} were verified against a separate high-resolution calculation that included Dopper shifting of lines. 
Therefore it's possible that the expansion limiting effect described above would not have a strong impact in cases where the emission is thermal due to a thick line forest.

The scheme above assumes that the thermalization time for the line resonance region is short relative to the dynamical time. This is often true for strong lines, as thermalization times can be quite short, typically $30 \ \rm{msec} \times \rho_{13}^{-1} \kappa_{4}^{-1}$, where $\rho\equiv\rho_{13}\times 10^{-13} \ \rm{g \ cm}^{-3}$ and $\kappa_\nu\equiv \kappa_{4} \times 10^{4}\ \rm cm^2 \ g^{-1}$. For a forest of weak lines, thermalization would proceed more slowly, but such lines would also be more moderately affected by the proposed upper limit. In addition, as $t_{exp}$ increases and weaker lines become affected by the expansion limit, the dynamical time grows as well.

\section{Summary and Discussion}
\label{sec: Summary Discussion}

In this letter we showed that the choice of line treatment in simulation can have an orders of magnitude effect on the introduced opacity, focusing on non-relativistic optically thick plasma.
In \S~\ref{sec: bound-free EOS} and fig. \ref{fig: bf opacity}, we compared the \textit{bound-free} opacity from our opacity table in \citet[][- \citetalias{morag_frequency_2023}]{morag_frequency_2023}, against the opacity given in the literature by \citet[][- \citetalias{blinnikov_comparative_1998}]{blinnikov_comparative_1998} , finding orders of magnitude difference. We concluded that the discrepancy in the tables in bound-free processes is likely due to deviations in the underlying plasma equation of state (EOS).
In \S~\ref{sec: bound-bound opacity} (figs. \ref{fig: bb opacity higher-res} and \ref{fig: bb opacity Blin vs ours}) we made a similar comparison, focusing on \textit{bound-bound} opacity. We were able to reproduce the opacities presented in \citetalias{blinnikov_comparative_1998} by employing the \citetalias{eastman_spectrum_1993} prescription (eqs. \ref{eq: Eastman Pinto 93} and \ref{eq: tau sobolev}). At the expansion time $t_{\rm exp}$ presented, this reproduced opacity was found to be orders of magnitude lower than the `static' average absorption opacity $\left<\kappa_\nu\right>_{\rm i}$ extracted from the high-resolution \citetalias{morag_frequency_2023} table. In \S~\ref{sec: alternative approach}, we proposed a modification to the average emission / absorption opacity that includes a limit on line-strength based on photon frequency shift during expansion. We show an example of this limit in fig. \S~\ref{fig:example opacity t_exp limit}.  We reiterate our formula modifying the Sobolev line strength $\kappa_l$, given by
\begin{equation}
\kappa_{l,exp}=\min[\kappa_l, (\rho c t_{exp})^{-1}]
\label{eq: line expansion limit}
\end{equation}

The different opacity choices can lead to important differences in the calculated SED. If the approximate static $\left<\kappa_\nu\right>_{\rm i}$ is used to calculate emissivity in radiative transfer calculations, the presence of a line forest at energies $h\nu\gtrsim3T$ can lead at peak energy frequencies to considerable reprocessing, even out to low scattering optical depths ($\tau<1$). The resulting SED at these frequencies would then often appear similar to a blackbody at the local temperature (if the plasma itself is near LTE). The effect of photoionization peaks in this case would then generally be negligible. On the other hand, if the \citetalias{eastman_spectrum_1993} approximation is used instead, photons formed inside at many scattering optical depths ($\tau \sim$100's-1000's) can undergo only limited reprocessing. The peak  photon energy would be representative of the temperatures at these depths, leading to a shift of the peak relative to a blackbody. In this case the sharp effect of photoionization opacity can become more observable as well.

The \citetalias{eastman_spectrum_1993} prescription is useful in the context of radiative transfer for calculating the photon mean-free-path, as the photon shifts in frequency across transition lines. It
is less useful for calculating the photon emission and absorption rates (emissivity) $c\rho\kappa_\nu$, as it ignores the maximum line strength (recall that the contribution from each line is capped at $(\nu/\Delta\nu)_{\rm i}(ct_{\rm exp})^{-1}\times 1$). The correct total photon production rate should be given by $\left<\kappa_\nu\right>_{\rm i}$ independent of velocity shear. This approach however is also not ideal, as the photons in a coarse multi-group calculation are assumed to be produced uniformly in frequency across a group, which is not the case physically. Meanwhile, a correct description of the photon \textit{absorption} is even more complex, as it depends on the local photon energy distribution $u_\nu$ that can be uneven at line frequency resolutions.

It is possible that the true spectrum results lie somewhere in between the two prescriptions. The physically correct emissivity must exceed \citetalias{eastman_spectrum_1993}, but it is also possible to overestimate the emission / absorption processes when using the average absorption opacity. This is a motivation for using the opacity cutoff formula in \S~\ref{sec: alternative approach}, which should yield an emissivity that lies in between the two extremes. We note however that it is possible for many cases that the modified formula will not have a strong effect relative to using $\langle \kappa_\nu \rangle_i$. The reason being that in the thermalized limit of strong absorption, there is weak sensitivity to the exact absorption amount. This insensitivity occurs since additional absorption, regardless of the details, only helps to further maintain thermal equilibrium. We also previously verified the \citetalias{morag_shock_2024} results that used $\left<\kappa_\nu\right>_{\rm i}$ for shock-cooling emission by showing agreement with a separate post-processing calculation. This high frequency resolution calculation resolved individual lines and included the effect of expansion opacity, providing a separate measure of validity.

An updated version of our frequency-dependent opacity table is now available in \citet{morag_frequency_2023}. The calculations for producing a high resolution table are the same as described in the previous version in \citetalias{morag_shock_2024}.
However, we add additional features for use in analysis and simulation.
These include functions that provide opacity $\kappa_R$ and $\left<\kappa_\nu\right>_{\rm i}$ averages for coarse frequency resolution MG simulations. The update includes approximations for the \citetalias{eastman_spectrum_1993} expansion opacity, including (eqs. \ref{eq: Eastman Pinto 93} and \ref{eq: tau sobolev}) for MG simulations. It also allows for direct broadening and shifting of the high-resolution table for arbitrary choices of $v/c$. Finally it includes the option of introducing the line expansion limit described in eq. \ref{eq: line expansion limit}.

\section*{Acknowledgements}
The author would like to thank Eli Waxman for his collaboration, as well as Ehud Nakar, Kyohei Kawaguchi, and Kenta Hotokezaka
for insightful discussion.

\section*{Data Availability}
Our opacity table code is available online for public use on \href{https://github.com/jon-morag/Freq_Dept_Opac_Table/}{github}.

% \clearpage
\bibliographystyle{mnras}
\bibliography{references}

\appendix

\section{Additional Comments on calculting opacities}

\begin{enumerate}
    \item 
It is reasonable in optically thick supersonic flows to calculate the scattering opacity as \citep[see also][]{mihalas_foundations_1999,castor_radiation_2007}
\begin{equation}
    \kappa_{\rm scat,i}\to\max(\kappa_{\rm R,i},\kappa_{\rm EP93,i}).
\end{equation}
This approach allows the introduction of line-wings into the calculation ($\kappa_{\rm R,i}$), as  they are not included in the Sobolev approximation upon which \citetalias{eastman_spectrum_1993} is based. Line wings can have a significant effect on the continuum.

\item
In the literature, lines interactions are often assumed to be either absorption or scattering events by adjusting a single parameter \citep[usually absorption - ][]{baron_non-local_1996,blinnikov_comparative_1998,hillier_time-dependent_2012,kozyreva_influence_2020}. Several works address this head-on by relaxing the assumption of LTE in the plasma and calculating specific relaxation rates for competing atomic processes, albeit for a subset of the species \citep[e.g.][]{lucy_monte_2002,lucy_monte_2003,kromer_artis_2021,pinto_physics_2000,dessart_numerical_2015}.
For other works that assume local thermal equllibrium of the plasma (LTE), we suggest that the frequency-dependent emission/absorption opacity may be modified as
\begin{equation}
    \kappa_{\nu,\rm abs}\to\min(\kappa_{\nu,\rm abs},R/\rho c),
\end{equation}
where the relaxation rate $R={\rm func}(l,\nu,\rho,T,\rm composition)$ can be estimated analytically -with some care-. In this suggested scheme, atomic transitions that emit and absorb faster than the relaxation rate will reprocess at most at the relaxation rate, and will behave as scattering events above this rate.

\item
\citetalias{morag_frequency_2023}  implements the \cite{hummer_equation_1988} micro-plasma suppression factor, which is given as a function of quantum number $n$ for H-like ions only. For mixtures that are not dominated by H, the user may choose to implement an approximate extension for non-H-like ions. Under the approximation that a sufficiently excited single electron sees a screened H-like ion, its effective H-like quantum number is then given by 
\begin{equation}
 n_{\rm eff}=Z_{\rm i}\sqrt{I_{\rm R}/\left( I_{\rm i}-E_{\rm s} \right)}.
\end{equation}
Here $Z_{\rm i}$ is the net screened nuclear charge number the excited electron sees, $I_{\rm R}=13.6$ eV, $I_{\rm i}>0$ is the ionization energy of the particular ionization state, and $E_{\rm s}>0$ is the energy of the state measured from the ionization ground state. This prescription suppresses multiple excited electron states (perhaps unnaturally) due to their higher energy.
\end{enumerate}

% Don't change these lines
\bsp	% typesetting comment
\label{lastpage}
\end{document}